\documentclass{aa}
\usepackage{graphicx}
\usepackage{txfonts}
\usepackage{natbib}
\usepackage{color}
\usepackage{multirow}
\usepackage{multicol}
\usepackage[colorlinks, allcolors=blue]{hyperref}
\newcommand{\HRI}{HRI$_\mathrm{EUV}$}

\begin{document}

\title{Spatial distributions of EUV brightenings in the quiet-Sun}

\author{C. J. Nelson$^{1}$, L. A. Hayes$^{1}$, D. M\"uller$^{1}$, S. Musset$^{1}$, N. Freij$^{2,3}$, F. Auch\`ere$^{4}$, R. Aznar Cuadrado$^{5}$, \\K. Barczynski$^{6,7}$, E. Buchlin$^{4}$, L. Harra$^{7,6}$, D. M. Long$^{8,9}$, S. Parenti$^{4}$, H. Peter$^{5}$, U. Sch\"uhle$^{5}$, \\P. Smith$^{10}$, L. Teriaca$^{5}$, C. Verbeeck$^{11}$, A. N. Zhukov$^{11,12}$, D. Berghmans$^{11}$}

\offprints{chris.nelson@esa.int}
\institute{$^{1}$European Space Agency (ESA), European Space Research and Technology Centre (ESTEC), Keplerlaan 1, 2201 AZ Noordwijk, The Netherlands\\
$^{2}$Lockheed Martin Solar and Astrophysics Laboratory, Palo Alto, CA, United States\\
$^{3}$Bay Area Environmental Research Institute, Moffett Field, CA, United States\\
$^{4}$Universit\'e Paris-Saclay, CNRS, Institut d’astrophysique spatiale, 91405, Orsay, France\\
$^{5}$Max Planck Institute for Solar System Research, Justus-von-Liebig-Weg 3, 37077 Göttingen, Germany\\
$^{6}$ETH-Zurich, Wolfgang-Pauli-Str. 27, 8093 Zurich, Switzerland\\
$^{7}$Physikalisch-Meteorologisches Observatorium Davos, World Radiation Center, 7260, Davos Dorf, Switzerland\\
$^{8}$School of Physical Sciences, Dublin City University, Glasnevin Campus, Dublin, D09 V209, Ireland\\
$^{9}$Astrophysics Research Centre, School of Mathematics and Physics, Queen's University Belfast, Belfast BT7 1NN, UK\\
$^{10}$UCL-Mullard Space Science Laboratory, Holmbury St. Mary, Dorking, Surrey, RH5 6NT, UK\\
$^{11}$Solar-Terrestrial Centre of Excellence – SIDC, Royal Observatory of Belgium, Ringlaan -3- Av. Circulaire, 1180 Brussels, Belgium\\
$^{12}$Skobeltsyn Institute of Nuclear Physics, Moscow State University, 119992 Moscow, Russia}

\date{}

\abstract
{The identification of large numbers of localised transient extreme ultraviolet (EUV) brightenings, with very small spatial scales, in the quiet-Sun corona has been one of the key early results from Solar Orbiter. However, much is still unknown about these events.}
{Here, we aim to better understand EUV brightenings by investigating their spatial distributions, specifically whether they occur co-spatial with specific line-of-sight magnetic field topologies in the photospheric network.}
{EUV brightenings are detected using an automated algorithm applied to a high-cadence ($3$ s) dataset sampled over $\sim30$ minutes on $8$ March 2022 by the Extreme Ultraviolet Imager's $17.4$ nm EUV High Resolution Imager (\HRI). Data from the Solar Dynamics Observatory's Helioseismic and Magnetic Imager (SDO/HMI) and Atmospheric Imaging Assembly (SDO/AIA) are used to provide context about the line-of-sight magnetic field and for alignment purposes, respectively.}
{We found a total of $5064$ EUV brightenings within this dataset that are directly comparable to events reported previously in the literature. These events occurred within around $0.015$-$0.020$ \%\ of pixels for any given frame. We compared eight different thresholds to split the EUV brightenings into four different categories related to the line-of-sight magnetic field. Using our preferred threshold, we found that $627$ EUV brightenings  ($12.4$ \%) occurred co-spatial with Strong Bipolar configurations and $967$ EUV brightenings ($19.1$ \%) occurred in Weak Field regions. Fewer than $10$ \%\ of EUV brightenings occurred co-spatial with Unipolar line-of-sight magnetic field no matter what threshold was used. Of the $627$ Strong Bipolar EUV Brightenings, $54$ were found to occur co-spatial with cancellation whilst $57$ occurred co-spatial with emergence.}
{EUV brightenings preferentially occur co-spatial with the strong line-of-sight magnetic field in the photospheric network. They do not, though, predominantly occur co-spatial with (cancelling) bi-poles.}

\authorrunning{Nelson et al.}
\titlerunning{Spatial distributions of EUV brightenings in the quiet-Sun}

\maketitle

\section{Introduction}
\label{Introduction}

Localised transient brightenings occur throughout the solar atmosphere from the lower photosphere to the upper corona (\citealt{Young18}). This diverse family of events typically have measured diameters which follow power-law distributions extending right down to the spatial resolutions of whichever instrument is used to observe them. This means that advances in instrumentation often lead to important discoveries in the field as smaller spatial scales and different thermal windows are sampled. One recent example of an instrument-linked step forward was the discovery of ubiquitous transient, localised EUV brightenings low down in the quiet-Sun corona by \citet{Berghmans21}. The authors used data from the Extreme Ultraviolet Imager's (EUI; \citealt{Rochus20}) $17.4$ nm EUV High-Resolution Imager (\HRI), on-board the European Space Agency (ESA) led Solar Orbiter (\citealt{Muller20}) spacecraft, to identify close to $1500$ EUV brightenings, with lifetimes up to $200$ s and lengths up to $4$ Mm, in a dataset spanning less than $5$ minutes. We note that these events occur on similar scales to some other localised brightenings detected in coronal imaging data (see, for example \citealt{Peter13, Barczynski17}) using the short time-series high-resolution datasets sampled by Hi-C (\citealt{Cirtain13}). 

\begin{figure*}
\includegraphics[width=0.99\textwidth]{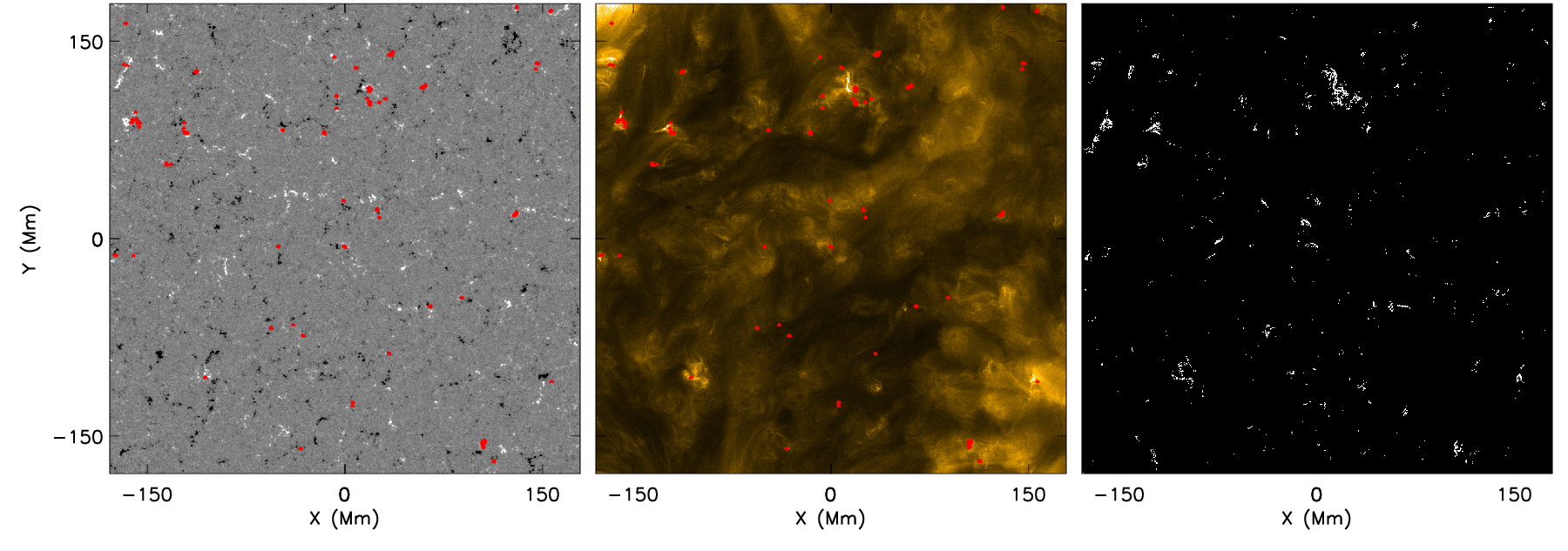}
\caption{Overview of the field-of-view (FOV) analysed in this article. Left panel: The line-of-sight magnetic field (where white is positive polarity and black is negative polarity) saturated at $\pm100$ G as measured by SDO/HMI at around $00$:$09$:$41$ UT. Middle panel: The same FOV as sampled co-temporally by the \HRI\ $17.4$ nm filter. The red contours in both these panels outline the EUV brightenings detected in this frame using the algorithm described in Sect.~\ref{Detection}. Right panel: A binary map plotting the locations of all pixels (white) identified to contain an EUV brightening meeting the criteria outlined in Sect.~\ref{Detection} in at least one frame within this dataset.}
\label{CFs_fig1}
\end{figure*}

Despite the recent discovery of EUV brightenings on such small scales, the community has been able to make rapid progress towards understanding these events over the last three years. \citet{Zhukov21} used the angular separation between EUI and the Solar Dynamics Observatory's Atmospheric Imaging Assembly (SDO/AIA; \citealt{Lemen12}) to conduct stereoscopy on these events, determining that EUV brightenings were typically formed at heights associated with the transition region or the lower corona (between 1 and 5 Mm). \citet{Panesar21}, \citet{Dolliou23}, and \citet{Dolliou24} observed that these events often occurred after a `cool' component detectable in various imaging and spectroscopic data, whilst \citet{Nelson23} studied the co-spatial signatures of EUV brightenings in data sampled by the Interface Region Imaging Spectrograph (IRIS; \citealt{DePontieu14}) finding that no typical response was present in \ion{Mg}{II} and \ion{Si}{IV} imaging data. Notably, Explosive Event (\citealt{Brueckner83}) spectra (broad, asymmetric intensity increases in transition region lines) were observed co-spatial with the single EUV brightening sampled by the IRIS spectrograph, observing plasma between $20-100$ kK. \citet{Alipour22} found $27$ \%\ of their automatically detected EUV brightenings occurred within larger-scale coronal bright points (presumably co-spatial with large photospheric network elements), hinting that several different types of EUV brightenings may be present in \HRI\ data. Similar complexity has been found by other authors such as \citet{Schwanitz23}.

In addition to this, numerous authors have investigated the relationship between EUV brightenings and the local magnetic field topology. \citet{Panesar21} studied the line-of-sight magnetic field structuring at the bases of EUV brightenings using the Helioseismic and Magnetic Imager (SDO/HMI; \citealt{Scherrer12}) finding that most of the $52$ events identified occurred close to polarity inversion lines and magnetic cancellation. \citet{Kahil22} conducted a similar study except using data sampled by the higher-resolution Polarimetric and Helioseismic Imager (PHI; \citealt{Solanki20}) and found bipoles co-spatial with $71$ \%\ of the $38$ EUV brightenings studied. Although both of these analyses returned important results, it should be noted that they analysed only a tiny fraction of EUV brightenings, given several thousand are present in any given dataset. On top of this, \citet{Barczynski22} compared the statistics of observed EUV brightenings with events detected in synthetic coronal images constructed from non-potential extrapolations of SDO/HMI line-of-sight magnetic field maps. Those authors concluded that the formation of EUV brightenings is consistent with energy release as the corona evolves between non-linear force-free states. \citet{Chen21} performed 3D radiation magnetohydrodynamics (MHD) simulations using the MURaM code with the coronal extension (\citealt{Vogler05, Rempel17}) and were able to simulate events with similar properties to observed EUV brightenings. The coronal magnetic field at those locations appeared to evolve in a manner consistent with component magnetic reconnection for the majority of events. 

In this article, we expand on these previous researches by conducting a large-scale statistical analysis of the relationship between more than $5000$ EUV brightenings and the line-of-sight magnetic field in the photosphere. Our work is set out as follows: In Sect.~\ref{Observations} we introduce the data studied here and the methods used to align and analyse them; in Sect.~\ref{Results} we present the results from our analysis; in Sect.~\ref{Discussion} we include a discussion of our results; before we outline our conclusions in Sect.~\ref{Conclusions}.

\begin{figure*}
\includegraphics[width=0.99\textwidth]{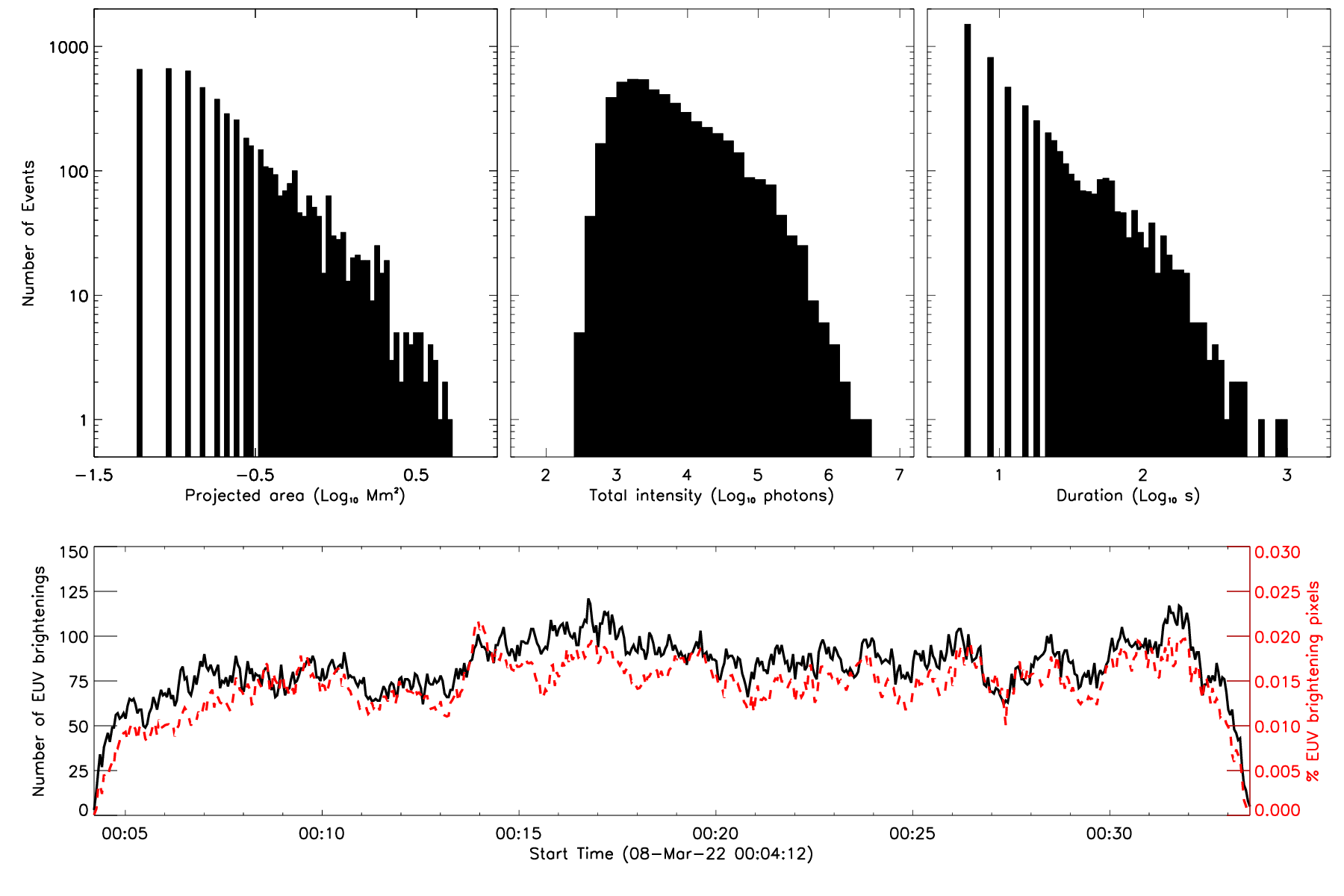}
\caption{Statistical properties of the $5064$ EUV brightenings that match the criteria outlined in Sect.~\ref{Detection} identified within this dataset by the automated detection algorithm. Top left panel: A histogram plotting the peak projected area of each EUV brightening (log-scaled along both the x- and y-axes) binned in the log-scale. Top middle panel: A histogram plotting the total intensity of each EUV brightening summed across space and time (log-scaled along the x- and y-axes) binned in the log-scale. Top right panel: A histogram plotting the duration of each EUV brightening (log-scaled along both the x- and y-axes) binned in the log-scale. Bottom panel: A time-series plotting the number of EUV brightenings (black solid line) and the percentage of pixels found to be within EUV brightenings (red dashed line) throughout this dataset. Note, both of these metrics tend to zero at each end of the time-series as we remove EUV brightenings found in the first and last frames from our sample (which reduces the number of events detected close to either end of the time-series).}
\label{CFs_fig2}
\end{figure*}

\section{Observations}
\label{Observations}

\subsection{Data products}

In this article, we analyse a region of the quiet-Sun sampled by \HRI\ between $00$:$04$:$12$ UT and $00$:$33$:$33$ UT (all times are corrected to the equivalent light arrival times at Earth) on $8$ March $2022$. During this time, Solar Orbiter was at a distance of approximately $0.49$ au from the Sun (leading to a time-delta of $251$ s between light arriving at the satellite and at Earth) and operating within its first Remote Sensing Window (RSW) of the nominal mission phase, conducting an instance of the `Nanoflares' Solar Orbiter Observing Plan (SOOP; \citealt{Zouganelis20}). A summary of EUI datasets availability from the first three RSWs can be found in \citet{Berghmans23}. On $8$ March $2022$, Solar Orbiter had an angular separation of around $3.4^\circ$ from Earth allowing good coordination with Earth-bound observational infrastructure. Here, we use level-2 calibrated \HRI\ data downloaded directly from the Solar Orbiter Archive\footnote{EUI data release 5.0: https://doi.org/10.24414/2qfw-tr95}. The plasma imaged by \HRI\ using the $17.4$ nm filter has a peak temperature response centred close to $10^{6}$ K, allowing us to gain insights into the dynamics of the lower solar corona. These $2048\times2048$ pixel$^2$ images have a pixel scale of $0.492$\arcsec\ and a cadence of $3$ s per frame on average, returning $588$ images within the time-period studied. The pixel scale of \HRI\ at 0.49 au is equivalent to approximately 174 km on the Sun. \citet{Berghmans23} (Fig. 24) estimated the full-width half-maximum of the \HRI\ point spread function to be 1.5 pixels in flight, implying the spatial resolution of this telescope can be defined by the Nyquist sampling limit of 2 pixels, or 348 km on the Sun. We note that this dataset has been studied previously by \citet{Nelson23} and \citet{Dolliou24}.

Additionally, we study co-spatial imaging and line-of-sight magnetic field maps collected by the SDO/AIA and SDO/HMI instruments, respectively. Specifically, we study narrow-band imaging sampled using the $17.1$ nm filter of SDO/AIA, which observes plasma with temperatures close to those observed by \HRI. During this time, $17.1$ nm images were collected by SDO/AIA with a cadence of $6$ s (with every $15$th frame having a cadence of $12$ s) allowing us to conduct a better alignment process. The line-of-sight magnetic field maps collected by the SDO/HMI instrument have post-reduction pixel scales of $0.6$\arcsec\ ($\sim435$ km), cadences of $45$ s, and a noise level of around $10$ G. The centre of the \HRI\ field-of-view (FOV) corresponds to helioprojective coordinates of ($-33$\arcsec, $121$\arcsec) as observed from SDO, whilst the FOV is around $356\times356$ Mm$^2$. In the left two panels of Fig.~\ref{CFs_fig1}, we plot the FOV studied here as observed by SDO/HMI and \HRI. The red contours over-laid on these panels outline the EUV brightenings detected, using an automated detection algorithm, in this \HRI\ frame that match the criteria defined in Sect.~\ref{Detection}. In general, the majority of EUV brightenings appear to be located close to strong line-of-sight magnetic field that maps the photospheric network. The white pixels within the binary map plotted in the right-most panel of Fig.~\ref{CFs_fig1} indicate the locations of all pixels found to contain an EUV brightening from our sample of $5064$ in at least one frame within this dataset. The correlation between EUV brightening locations and the local line-of-sight magnetic field will be investigated in more detail in Sect.~\ref{Results}.

\begin{figure*}
\includegraphics[width=0.99\textwidth]{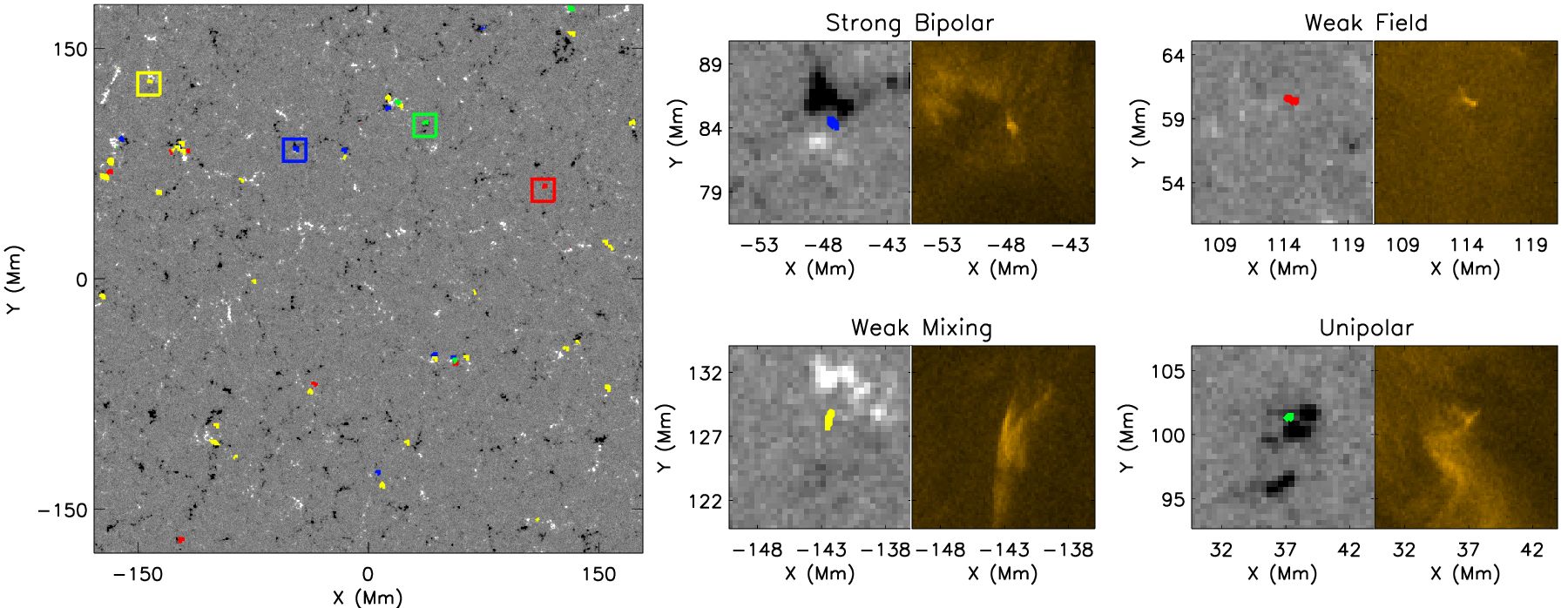}
\caption{Relationship between EUV brightenings and the line-of-sight magnetic field in the quiet-Sun. Left panel: The background image plots a line-of-sight magnetic field map measured by SDO/HMI at around $00$:$29$:$42$ UT, saturated at $\pm100$ G. The over-laid contours outline EUV brightenings with the different colours representing different line-of-sight magnetic field configurations in the photosphere, measured within a $5\times5$ pixel$^2$ SDO/HMI FOV centred on the middle of the EUV brightening. Blue contours indicate EUV brightenings where Strong Bipolar field (both polarities with maxima above $|20|$ G) is measured co-spatially. Red contours indicate EUV brightenings where only Weak Field (less than $|20|$ G) is measured co-spatially. Green contours indicate EUV brightenings where Unipolar (e.g. only positive or only negative polarities measured, with the present polarity having a maximum field strength above $|20|$ G) field is measured co-spatially. The yellow contours indicate EUV brightenings where one strong polarity (greater than $|20|$ G) and one weak polarity (less than $|20|$ G) are measured co-spatially. The four coloured boxes outline the representative EUV brightenings that are plotted in the right hand panels. Right panels: SDO/HMI line-of-sight magnetic field maps (left panels) and \HRI\ $17.4$ nm intensity maps (right panels) of EUV brightenings (outlined by coloured contours) where Strong Bipolar (top left), Weak Field (top right), Weak Mixing (bottom left), and Unipolar (bottom right) line-of-sight magnetic field is measured co-spatially. A movie associated with this figure is available with the online version of this article.}
\label{CFs_fig3}
\end{figure*}

\subsection{EUV brightening detection}
\label{Detection}

In order to detect EUV brightenings in a manner consistent with the previous literature, the algorithm developed by \citet{Berghmans21} was applied to Carrington remapped grids of the field-of-view (FOV) through time. This remapping increased the image dimensions to $2172\times2172$ pixel$^2$ and included steps to account for instrument jitter (see Sect.~\ref{Alignment}). We include a brief overview of the algorithm here for completness. Following the methodology described in~\citep{Starck1994, Murtagh1995} for denoising images, the detected events are defined as the significant coefficients in the first two spatial scales of an {\it \`a trous} wavelet decomposition of the input images, using a $B_3$ spline scaling function. The amplitude of the coefficients at each scale is compared pixel-per-pixel to that expected in the presence of noise only. The confidence level is set to be five times the spatially variable standard deviation computed from the combination of photon shot noise and detector read noise. The detection is performed frame-by-frame, resulting in a space-time binary cube. The six-connected voxels are clustered into detected space-time events. Overall, a total of $13864$ EUV brightenings were detected within this dataset with a birthrate of $5.6\times10^{-17}$ m$^{-2}$ s$^{-1}$ (comparable to the birthrate of $3.7\times10^{-17}$ m$^{-2}$ s$^{-1}$ for these events calculated by \citealt{Berghmans21}).

\begin{table*}[!hbt]
\centering
\caption{Statistics of EUV brightenings related to the line-of-sight magnetic field configuration for eight different detection thresholds.}
\begin{tabular}{ c  c  c  c  c  c  c  c  c  c  c  c  c  c  c  c  c  c  c }
\hline
\multirow{2}{*}{\bf{B}$_\mathrm{th}$} & HMI & & \multicolumn{3}{c }{\color{blue}\bf{Strong Bipolar}} & & \multicolumn{3}{c }{\color{yellow}\bf{Weak Mixing}} & & \multicolumn{3}{c }{\color{green}\bf{Unipolar}} & & \multicolumn{3}{c }{\color{red}\bf{Weak Field}}\\
& FOV & &  Events & $\overline{\mathrm{area}}$ & $\overline{\mathrm{dur}}$ & & Events & $\overline{\mathrm{area}}$ & $\overline{\mathrm{dur}}$ & & Events & $\overline{\mathrm{area}}$ & $\overline{\mathrm{dur}}$ & & Events & $\overline{\mathrm{area}}$ & $\overline{\mathrm{dur}}$ \\
G & pixel$^2$ & & & Mm$^2$ & s & & & Mm$^2$ & s & & & Mm$^2$ & s & & & Mm$^2$ & s \\
\hline
20 & 5x5 & & 627 & 0.38 & 27 & & 3002 & 0.31 & 25 & & 468 & 0.25 & 27 & & 967 & 0.32 & 25 \\
30 & 5x5 & & 260 & 0.43 & 31 & & 2339 & 0.30 & 25 & & 468 & 0.25 & 27 & & 1997 & 0.33 & 25 \\
40 & 5x5 & & 159 & 0.45 & 29 & & 1912 & 0.30 & 25 & & 463 & 0.25 & 27 & & 2530 & 0.33 & 25 \\
50 & 5x5 & & 104 & 0.48 & 34 & & 1630 & 0.30 & 25 & & 457 & 0.25 & 27 & & 2873 & 0.32 & 25 \\
\hline
20 & 9x9 & & 2379 & 0.35 & 28 & & 2517 & 0.27 & 24 & & 22 & 0.21 & 32 & & 146 & 0.36 & 27 \\
30 & 9x9 & & 1141 & 0.40 & 31 & & 3114 & 0.28 & 24 & & 22 & 0.21 & 32 & & 787 & 0.31 & 25 \\
40 & 9x9 & & 763 & 0.40 & 30 & & 2972 & 0.29 & 25 & & 22 & 0.21 & 32 & & 1307 & 0.31 & 24 \\
50 & 9x9 & & 555 & 0.39 & 31 & & 2859 & 0.30 & 26 & & 22 & 0.21 & 32 & & 1628 & 0.31 & 24 \\ 
\hline
\end{tabular}

\tablefoot{The left two columns define the absolute line-of-sight magnetic field strength used as the cut-off between `Strong' and `Weak' field ({\bf{B}}$_\mathrm{th}$ in Sect.~\ref{Magnetic}) and the SDO/HMI FOV around each EUV brightening used to identify whether strong magnetic field was present ($m\times{m}$ in Sect.~\ref{Magnetic}), respectively. The number, average area, and average duration of EUV brightenings matching each of the four categories for the given threshold values are included in the remaining columns.}
\label{Tab_overview}
\end{table*}

Although the algorithm used here returns single pixel events, we only study those EUV brightenings that have areas equal to or above two pixels and durations equal to or above two frames. Additionally, we remove any EUV brightenings that are present in the first or last frame in order that we can study the entire evolution of each event. Applying these criteria returned a total of $5064$ events to be analysed here. In the top three panels of Fig.~\ref{CFs_fig2}, we plot histograms of the area (left), total intensity summed across both space and time (middle), and lifetime (right) of these $5064$ events. These histograms are qualitatively similar to those plotted by \citet{Berghmans21}, with the main differences being the detection of both larger (up to $5.2$ Mm$^{-2}$) and longer-lived (up to $945$ s) events. The detection of these longer-lived events is unsurprising as we study a longer time-series than those authors. In the bottom panel of Fig.~\ref{CFs_fig2}, we plot time-series of the number of EUV brightenings per frame (black solid line) and the percentage of pixels that host an EUV brightening (red dashed line). This plot indicates that each frame contains around $100$ EUV brightenings that, combined, make up around $0.015$-$0.020$ \%\ of pixels (similar to the coverage of IRIS bursts in active regions calculated by \citealt{Kleint22}). Note, that both of these metrics tend to zero at either end of the time-series due to our removal of EUV brightening events detected in the first and last frames.

\subsection{Alignment}
\label{Alignment}

Alignment between different instruments was conducted in multiple steps. Firstly, we remapped the EUV brightening binary maps returned by the algorithm in Carrington grids back to the original pixel scales using the $euvi\_heliographic.pro$ IDL routine. We note that single pixel events are possible within these new binary maps due to the rescaling. Secondly, we removed internal jitter from the \HRI\ time-series using the $tr\_get\_disp.pro$ IDL routine. Thirdly, we identified the SDO/AIA and SDO/HMI frames that were temporally closest to the first \HRI\ image, before using the $euvi\_heliographic.pro$ routine to identify the pixels in those full-disk images that were spatially closest to the centre of the \HRI\ data in Carrington coordinates (according to the header information). Fourthly, a FOV equivalent to the \HRI\ FOV was extracted from the SDO/AIA and SDO/HMI maps and tracked throughout the time-series using single pixel shifts as appropriate. Fifthly, these alignments were checked by creating movies of the SDO/AIA $17.1$ nm data with \HRI\ contours overlaid. Any obvious off-sets were corrected using manual pixels shifts (typically less than $2$ \HRI\ pixels). Although these alignments produce good results at the centre of the \HRI\ FOV, the different lines-of-sight of the satellites mean that this simple alignment has increasing errors as one moves towards the edge of the FOV. Sixthly, therefore, we constructed a binary map that identified all pixels within the \HRI\ data that contained an EUV brightening, before conducting a one-to-one mapping of this binary map onto the SDO/AIA and SDO/HMI FOV using the $euvi\_heliographic.pro$ routine. Any pixel shifts identified at the centre of the FOV were applied throughout.

\section{Results}
\label{Results}

\subsection{Magnetic structuring co-spatial to EUV brightenings}
\label{Magnetic}

After the detection of EUV brightenings and the alignment of the different datasets was completed, we investigated the relationship between these events and the co-spatial line-of-sight magnetic field topologies. To do this, we found the pixel and frame in the SDO/HMI time-series that corresponded to the centre, in both space and time, of each of the $5064$ EUV brightenings studied here. Next, small sub-FOVs were extracted from around these central points, with sizes of $m\times{m}$ pixel$^2$, before the line-of-sight magnetic field topology within this sub-FOV was sorted into one of four distinct categories. These categories were: Strong Bipolar, where both positive and negative polarity field have absolute maxima within the sub-FOV above an absolute threshold of {\bf{B}}$_\mathrm{th}$; Weak Mixing, where both polarities are present within the sub-FOV but only one has an absolute maximum above {\bf{B}}$_\mathrm{th}$; Unipolar, where only one polarity of magnetic field is present within the sub-FOV and this polarity has an absolute maximum above {\bf{B}}$_\mathrm{th}$; and Weak Field, where the absolute maxima of the local field is below {\bf{B}}$_\mathrm{th}$ within the sub-FOV. All references to `Strong' and `Weak', of course, refer to the line-of-sight magnetic field strength alone. Here, we conduct a parameter study using $m$ values of $5$ and $9$, and {\bf{B}}$_\mathrm{th}$ values of $20$ G, $30$ G, $40$ G, and $50$ G, which together provide eight thresholds to investigate. 

These thresholds are in some ways arbitrarily selected, however, they can also be justified through reasoning. For example, $m$ values of $5$ and $9$ return SDO/HMI FOVs of $2.2$ Mm$^2$ and $3.9$ Mm$^2$, respectively. These FOVs are large enough to fully encompass the vast majority of EUV brightenings, whilst also accounting for some errors in the alignment between instruments. Any smaller or larger FOVs would likely introduce additional noise into the categorisation. Additionally, {\bf{B}}$_\mathrm{th}$ values of $20$ G, $30$ G, $40$ G, and $50$ G provide a good range of magnetic field values to test against in the quiet-Sun, where weak line-of-sight magnetic field dominates. One could, potentially, also consider threshold values as low as the SDO/HMI noise level ($10$ G), although this is undesirable for a number of reasons. Perhaps most importantly, around $54$ \%\ of SDO/HMI pixels would be classified as `Strong Bipoles' at this threshold level implying that most EUV brightenings would be classified into the `Strong Bipoles' category when this was not actually the case visually. We note that we do not consider the EUV brightenings within the Weak Mixing category to be co-spatial to bipoles in our work. The main reason is that the Weak Mixing category contains events where one polarity may have a peak magnetic field strength below the SDO/HMI noise level meaning one cannot confidently state that both polarities are present at all. Overall, we are aware that no categorisation method is entirely perfect, but we are confident that our parameter study will allow us to obtain upper limits for the number of events co-spatial to each different magnetic field topology in the photosphere.

\begin{figure*}[!hbt]
\includegraphics[width=0.99\textwidth,trim={0 4.2cm 0 0}]{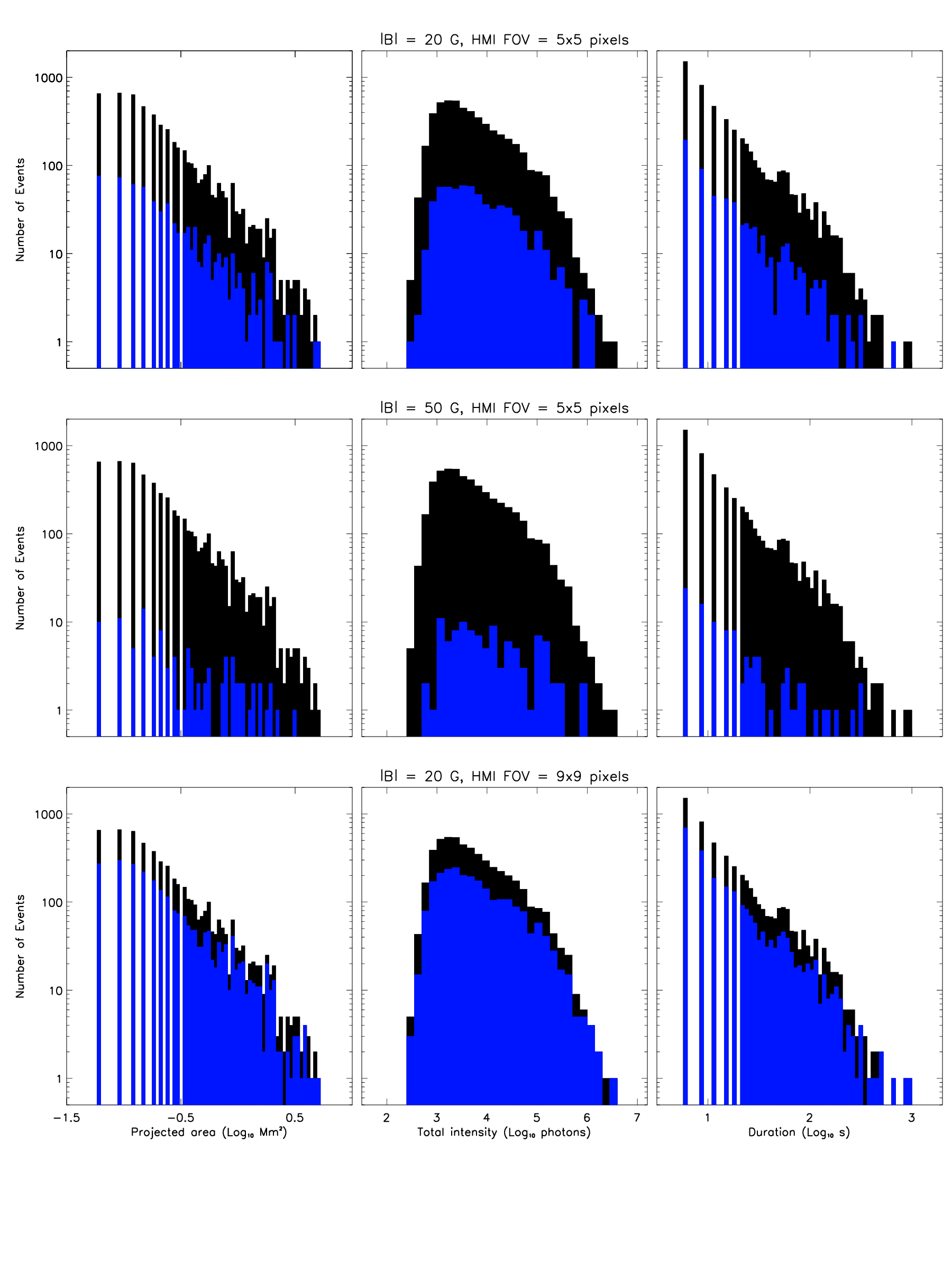}
\caption{These plots are the same as the top row of Fig.~\ref{CFs_fig2} but with the distributions of Strong Bipolar EUV brightenings overlaid (blue bars) for three different $m$ and {\bf{B}}$_\mathrm{th}$ pairings. Top row: Statistical distributions for our base pairing of $m=5$ and {\bf{B}}$_\mathrm{th}=20$ G. Middle row: Statistical distributions for our strictest pairing of $m=5$ and {\bf{B}}$_\mathrm{th}=50$ G. Bottom row: Statistical distributions for our loosest pairing of $m=9$ and {\bf{B}}$_\mathrm{th}=20$ G.}
\label{CFs_fig4}
\end{figure*} 

In Fig.~\ref{CFs_fig3}, we plot an example of this categorisation for an $m$ value of $5$ and a {\bf{B}}$_\mathrm{th}$ value of $20$ G. In the remainder of this article, we use these $m$ and {\bf{B}}$_\mathrm{th}$ values as our base against which all other values can be compared. In the left panel of Fig.~\ref{CFs_fig3}, we plot an SDO/HMI line-of-sight magnetic field map and overlay contours that outline EUV brightenings. The colours of the contours indicate which of the four categories of line-of-sight magnetic field configuration that respective EUV brightening is contained within. Namely, blue indicates Strong Bipolar, red indicates Weak Field, yellow indicates Weak Mixing, and green indicates Unipolar. The four similarly coloured boxes outline the regions for which both SDO/HMI line-of-sight magnetic field and \HRI\ $17.4$ nm maps are plotted in the right panels of Fig.~\ref{CFs_fig3} in order to demonstrate our method. Note, the regions plotted in the right panels of Fig.~\ref{CFs_fig3} are much larger than the sub-FOVs that are used to categorise the EUV brightenings, in order to allow the reader to better contextualise the local line-of-sight magnetic field topologies. Each of these regions contains an example EUV brightening as outlined by the contours. As these four example EUV brightenings are not representative of all events, we should be careful not to over-interpret the differences in the \HRI\ $17.4$ nm images; however, it is interesting to note that the EUV brightenings detected in the Strong Bipolar and Weak Field regions are qualitatively similar despite the very different line-of-sight magentic field configuration.

In Table~\ref{Tab_overview}, we present the statistics obtained using our categorisation method for each of the eight threshold sets studied. For our base values ($m=5$ and {\bf{B}}$_\mathrm{th}=20$ G), we find that $12.4$ \%\ ($627$) of EUV brightenings occur co-spatial with strong bipoles, $59.3$ \%\ ($3002$) of EUV brightenings occur co-spatial with Weak Mixing, $9.2$ \%\ ($468$) of EUV brightenings occur co-spatial with Unipolar field, and $19.1$ \%\ ($967$) of EUV brightenings occur co-spatial with Weak Field in the line-of-sight component. Within the entire SDO/HMI dataset, $46.2$ \%\ of pixels match the criteria of the Weak Field category suggesting, as the percentage of events co-spatial with Weak Field is much lower than this, that EUV brightenings form preferentially close to the strong (absolute values above $20$ G) line-of-sight magnetic field that comprises the photospheric network. Additionally, only $4.7$ \%\ and $1.1$ \%\ of pixels match the criteria for the Strong Bipolar and Unipolar categories, well below the percentage of EUV brightenings detected co-spatial with such line-of-sight magnetic field configurations. 

As expected, if we maintain our value of $m$ but increase our value of {\bf{B}}$_\mathrm{th}$ to $50$ G, the percentage of Strong Bipolar EUV brightenings detected decreases to $2.1$ \%, whilst the percentage of Weak Field EUV brightenings detected increases to $56.7$ \%. If, instead, we increase our value of $m$ to $9$ and maintain our value of {\bf{B}}$_\mathrm{th}$ at $20$ G, then the number of Strong Bipolar EUV brightenings increases to $47$ \%\, whilst the number of Unipolar and Weak Field EUV brightenings decrease to $0.4$ \%\ and $2.9$ \%, respectively. Given we consider values of $m=9$ and {\bf{B}}$_\mathrm{th}=20$ G to be overly generous for detecting true bipoles (due to the possibility of including `bipoles' that are separated by nearly $5$ Mm), we propose that $47$ \%\ is, therefore, an upper bound for the percentage of EUV brightenings that occur co-spatial with strong bipoles. We also propose that between $43.4$ \%\ and $97.1$ \%\ of EUV brightenings occur co-spatial with strong photospheric network elements (calculated using the number of EUV brightenings found to occur co-spatial with Weak Field using our loosest and strictest criteria), with a potentially `true' value being closer to $80.9$ \%\ (using our base criteria).

\begin{figure*}
\includegraphics[width=0.99\textwidth]{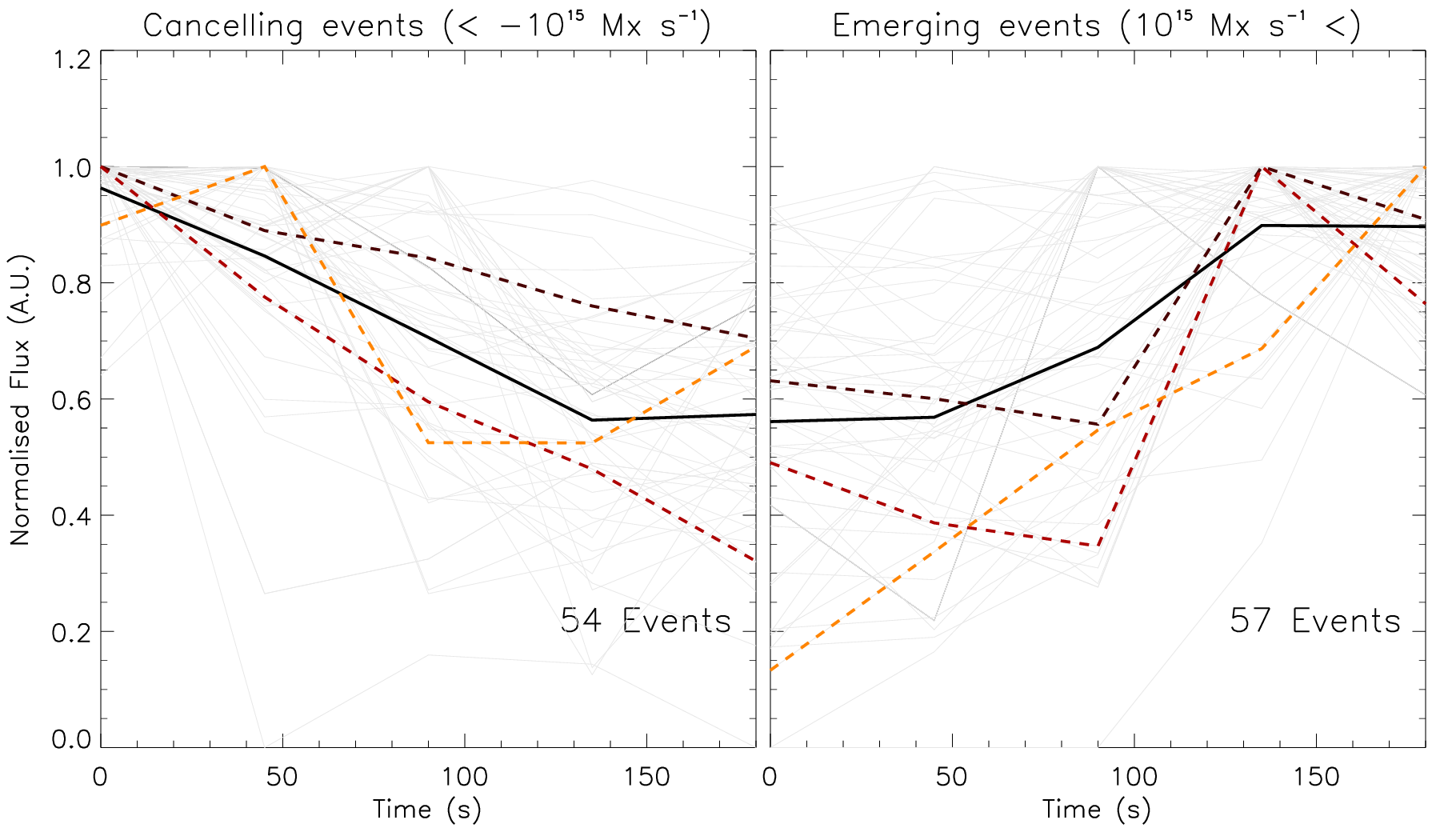}
\caption{Normalised evolution of the line-of-sight magnetic flux through time for the $111$ (out of $627$) Strong Bipolar EUV brightenings identified using our base pairing ($m=5$ and {\bf{B}}$_\mathrm{th}=20$ G) that display strong (thresholds of $\pm10^{15}$ Mx s$^{-1}$) evidence of cancellation ($54$) or emergence ($57$). Left panel: The normalised evolution of the line-of-sight magnetic flux within the SDO/HMI sub-FOV used to categorise EUV brightenings for the $54$ events found to have a cancellation rate below $-10^{15}$ Mx s$^{-1}$ (light grey lines). The overlaid dashed lines plot three example time-series. The solid black line plots the normalised flux at each time-step averaged over all $54$ events. Right panel: The normalised evolution of the line-of-sight magnetic flux within the SDO/HMI sub-FOV used to categorise EUV brightenings for the $57$ events found to have a emergence rate above $10^{15}$ Mx s$^{-1}$ (light grey lines). Once again, the dashed lines plot three example time-series and the black line plots the average time-series.}
\label{CFs_fig5}
\end{figure*}

Comparing the average areas and lifetimes of EUV brightenings within the four different categories returns additional interesting results. For $m$ values of $5$ (i.e., a $5\times5$ pixel$^2$ box around the EUV brightening in SDO/HMI data), we find that Strong Bipolar EUV brightenings have the biggest average areas for all {\bf{B}}$_\mathrm{th}$ values, with this metric increasing with {\bf{B}}$_\mathrm{th}$. Essentially, EUV brightenings are found to be larger, on average, when larger values are used as the cut-off between weak and strong field. Strong Bipolar EUV brightenings also have the longest (or equal longest for {\bf{B}}$_\mathrm{th}=20$ G) average lifetimes for all {\bf{B}}$_\mathrm{th}$ values when $m=5$. We find similar results for values of $m=9$, if the values calculated for the $22$ Unipolar events are excluded (due to their small sample size) during our comparison. 

In Fig.~\ref{CFs_fig4}, we plot the distributions of Strong Bipolar EUV brightening area (left column), intensity (middle column), and lifetime (right column) for three different pairing of $m$ and {\bf{B}}$_\mathrm{th}$. The top row plots distributions for our base values for $m$ ($5$) and {\bf{B}}$_\mathrm{th}$ ($20$ G). Qualitatively, the Strong Bipolar distributions (blue bars) are similar to the overall distributions of EUV brightenings. The middle row plots the distributions for our strictest pairing of $m$ ($5$) and {\bf{B}}$_\mathrm{th}$ ($50$ G). In this case, the Strong Bipolar events displaying a lower log-log gradient in the area and lifetime plots than the overall distribution (explaining why the average values for area and lifetime are highest for this category and this pairing of $m$ and {\bf{B}}$_\mathrm{th}$). We note, that only a small percentage ($2.1$ \%) of events are identified as being Strong Bipolar with this pairing of $m$ and {\bf{B}}$_\mathrm{th}$, though, meaning that a longer dataset should be studied in the future to confirm these potential differences in distributions. In the bottom row, we plot these distributions for our loosest pairing of $m$ ($9$) and {\bf{B}}$_\mathrm{th}$ ($20$ G). Here, once again, the distributions for the Strong Bipolar EUV brightenings appear to be similar to the distributions for all EUV brightenings. Overall, it appears that Strong Bipolar EUV brightenings are on average slightly larger and longer-lived than other EUV brightenings, with these differences becoming more pronounced when stricter conditions for defining a strong bipole are applied. However, importantly, not all large and long-lived EUV brightenings occur co-spatial with strong bipoles. A similar plot displaying the distributions of the EUV brightenings identified within the other categories is included in Fig.~\ref{CFs_fig1A}.

\begin{figure*}
\includegraphics[width=0.99\textwidth]{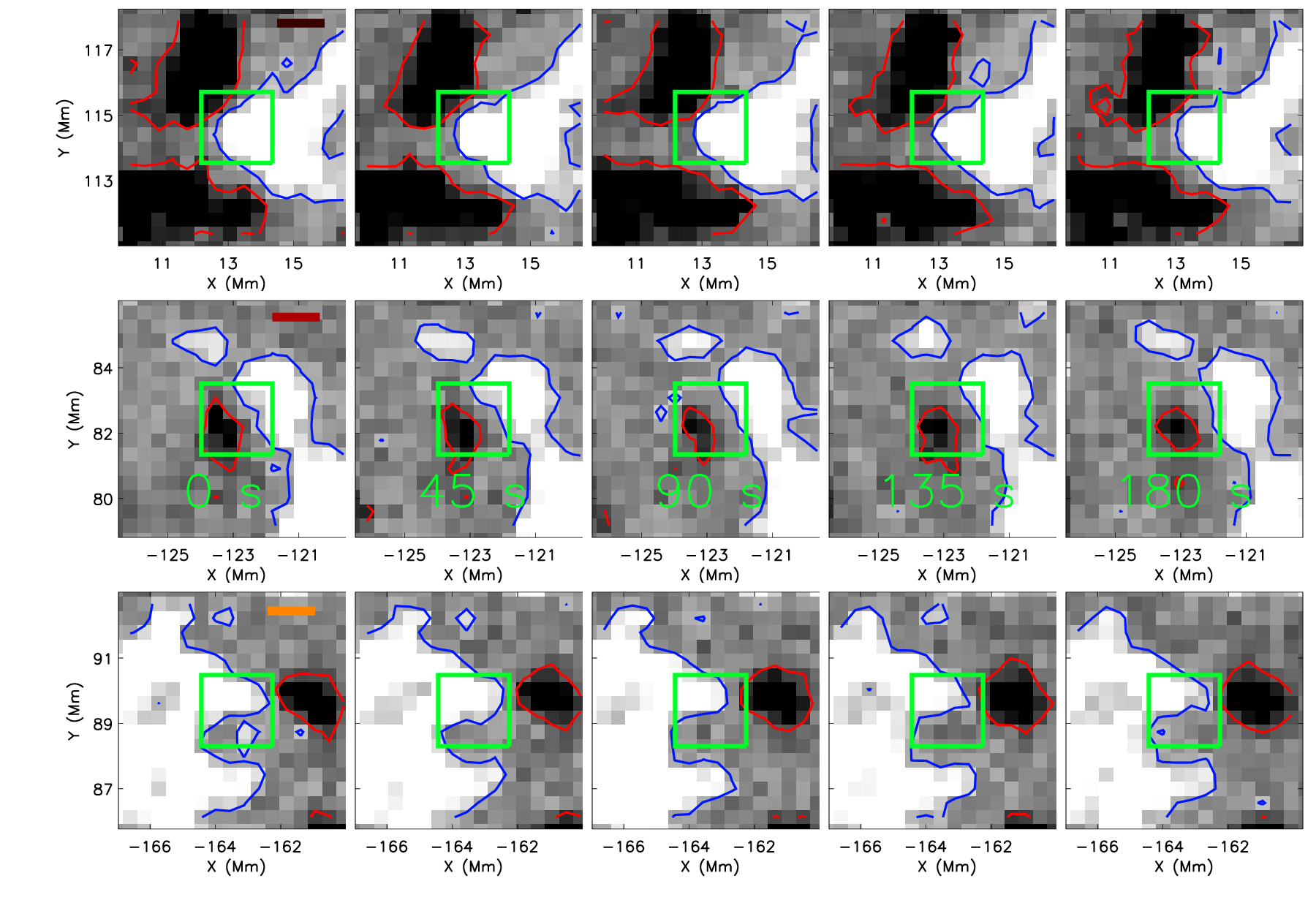}
\caption{Evolution of the SDO/HMI line-of-sight magnetic field during the occurrence of the three Strong Bipolar cancelling EUV brightenings plotted with the dashed lines in the left panel of Fig.~\ref{CFs_fig5}. The over-laid blue and red contours outline the regions where positive and negative line-of-sight magnetic field above an absolute strength of $20$ G, respectively, is measured. The green boxes outline the FOV used to categorise the events and to construct the normalised flux plots. The time between columns is $45$ s. Top row: Time-series for the darkest line (corresponding to the region where numerous cancelling and emerging EUV brightenings were detected). Middle row: Time-series for the red line. Bottom row: Time-series for the orange line. Short lines are included in the first row to indicate this more clearly on the figure.}
\label{CFs_fig6}
\end{figure*}

\subsection{Evolution of the line-of-sight magnetic field}

To continue our research into EUV brightenings, we also investigated whether the local magnetic field evolved in a specific manner (e.g. exhibiting cancellation or emergence) during their occurrence. Here, we limit our analysis to the $627$ EUV brightenings that were categorised as Strong Bipolar events using our base pairing ($m=5$ and {\bf{B}}$_\mathrm{th}=20$ G). For each event, we tracked the sub-FOV used to categorise the event across five SDO/HMI frames ($180$ s), temporally centred on the midpoint of the lifetime of that specific EUV brightening. The summed magnetic flux (only considering pixels with an absolute magnetic field strength above $20$ G) within the sub-FOV was then calculated for each frame, before a linear fit was applied to the time-series generated for each EUV brightening. The gradient of this linear fit was used to infer whether any evolution of the local line-of-sight magnetic flux was evident. For these $627$ events, $54$ displayed strong evidence of cancellation (rates $<-10^{15}$ Mx s$^{-1}$) and $57$ displayed strong evidence of emergence ($10^{15}$ Mx s$^{-1}<$ rates). The remaining $516$ events were split between weak cancellation ($-10^{15}$~Mx~s$^{-1}<$ rates $<-10^{14}$~Mx~s$^{-1}$; $186$ events), weak emergence ($10^{14}$~Mx~s$^{-1}<$ rates $<10^{15}$~Mx~s$^{-1}$; $157$ events), and no measurable change ($-10^{14}$~Mx~s$^{-1}<$ rates $<10^{14}$~Mx~s$^{-1}$; $173$ events).  If, instead, we consider all $5064$ events, we find similar results with $418$ ($8.3$ \%) EUV brightenings occuring co-spatial with increases in the line-of-sight magnetic flux (with rates above $10^{15}$ Mx s$^{-1}$) and $347$ ($6.9$ \%) EUV brightenings occuring co-spatial with reductions in the line-of-sight magnetic flux (with rates below $-10^{15}$ Mx s$^{-1}$). Clearly, no standard trend is observed in terms of line-of-sight magnetic field evolution during the occurrence of EUV brightenings.

\begin{figure*}
\includegraphics[width=0.99\textwidth]{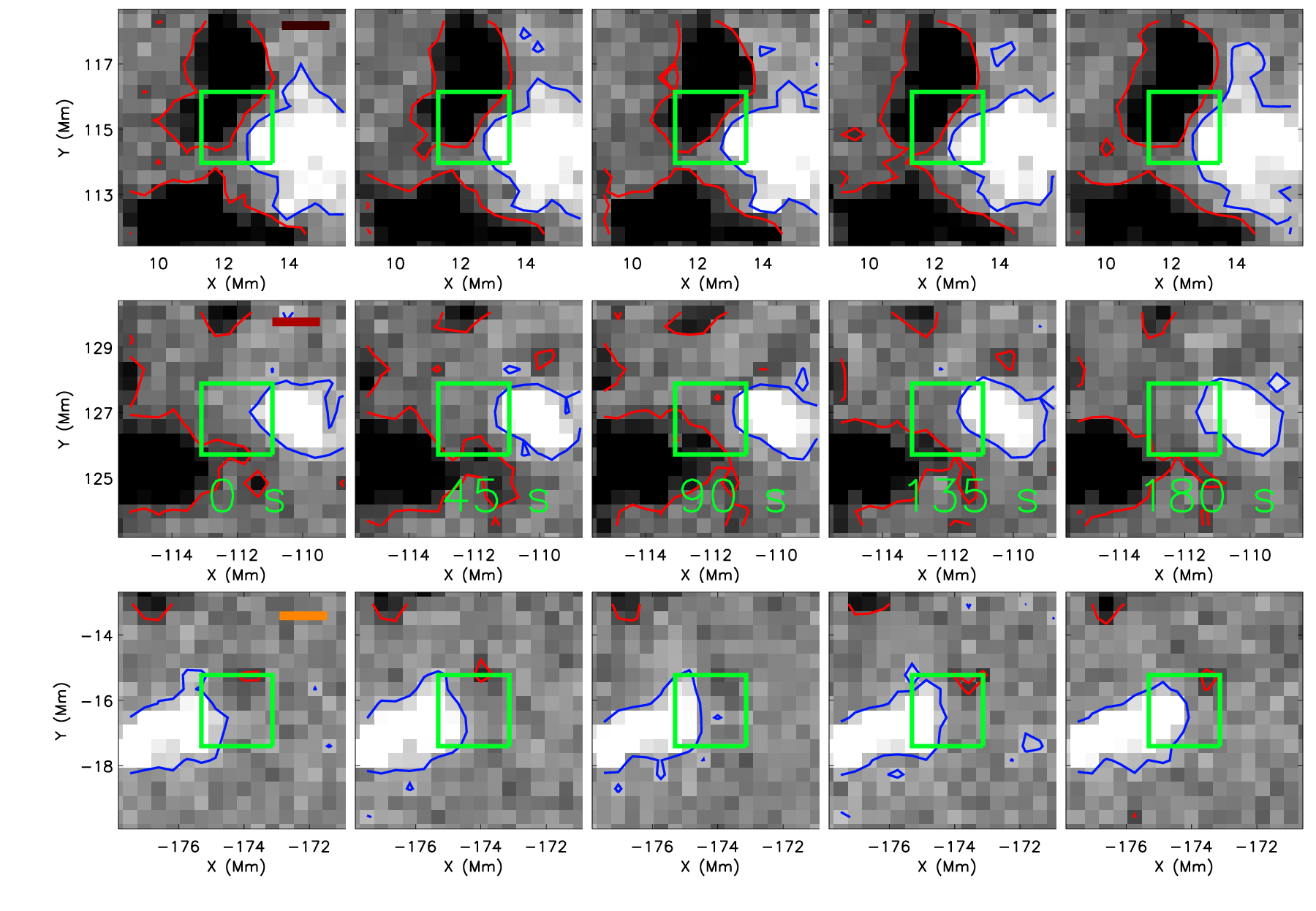}
\caption{These plots are the same as Fig.~\ref{CFs_fig6} but for the three Strong Bipolar emerging EUV brightenings plotted with the dashed lines in the right panel of Fig.~\ref{CFs_fig5}. Note, the top row plots an almost identical FOV to the top row of Fig.~\ref{CFs_fig6}.}
\label{CFs_fig7}
\end{figure*}

In Fig.~\ref{CFs_fig5}, we plot the evolution of the normalised flux for each of the Strong Bipolar EUV brightenings found to occur co-spatial with strong cancellation (left panel) and strong emergence (right panel). The light grey lines plot the evolution of the normalised flux for each individual EUV brightening, the coloured dashed lines plot three representative examples of the evolution of the normalised flux, whilst the solid black lines plot the average normalised flux at each time-step (calculated from the entire sample plotted). Interestingly, $10$ (out of $54$) of the Strong Bipolar EUV brightenings that display evidence of cancellation and $18$ (out of $57$) of the Strong Bipolar EUV brightenings that display evidence of emergence all come from the same small bipolar region, with central positions within a $9\times9$ pixel$^{2}$ SDO/HMI box at the centre of a bipolar network region (positioned at around $x_\mathrm{c}$=$13$\arcsec, $y_\mathrm{c}$=$115$\arcsec). The darkest dashed lines over-laid on Fig.~\ref{CFs_fig5} plot the normalised flux calculated co-spatial with one apparently cancelling (left) and one apparently emerging (right) Strong Bipolar EUV brightening from this region. If we were to consider all $5064$ EUV brightenings detected, then $60$ ($1.2$ \%) are found to be centred within the same $9\times9$ pixel$^{2}$ SDO/HMI box. We note that $620$ ($98.9$ \%) out of the $627$ Strong Bipolar EUV brightenings that were studied here have lifetimes below $180$ s indicating that our analysis should capture the relevant dynamics within the SDO/HMI FOV.

In Fig.~\ref{CFs_fig6}, we plot time-series of the measured SDO/HMI line-of-sight magnetic field for the three examples of Strong Bipolar cancelling EUV brightenings indicated by the dashed lines overlaid on the left panel of Fig.~\ref{CFs_fig5}. The top row plots the time-series for the darkest dashed line in the left panel of Fig.~\ref{CFs_fig5} (corresponding to the bipolar photospheric network region where many Strong Bipolar cancelling and emerging EUV brightenings were detected, as discussed in the previous paragraph), the middle row plots the time-series corresponding to the red dashed line, whilst the bottom row plots the time-series corresponding to the orange dashed line. The blue and red contours outline regions of positive and negative magnetic field (thresholds of $\pm20$ G), whilst the green boxes indicate the FOV used to categorise the events and to calculate the flux in the region at each time-step. The cadence between columns is $45$ s. No clear cancellation is apparent visually in the top row; however, some visible evidence of cancellation is apparent in the middle (negative polarity) and bottom (positive polarity) rows. In Fig.~\ref{CFs_fig7}, we plot the equivalent but for the three Strong Bipolar emerging EUV brightenings indicated by the dashed lines in the right panel of Fig.~\ref{CFs_fig5}. We note that the top row of Fig.~\ref{CFs_fig7} plots almost the same FOV as the top row of Fig.~\ref{CFs_fig6}, but for a different time-period. As with the cancelling events, no clear visual evidence of systematic changes in the line-of-sight magnetic field configuration are apparent in these FOVs. Overall, we find that EUV brightenings can occur co-spatial with both cancellation and emergence in line-of-sight magnetic field maps, but the presence of such magnetic field dynamics does not appear to be a necessary condition for EUV brightening formation.

\section{Discussion}
\label{Discussion}

\subsection{Links between EUV brightenings and the magnetic field}

Previous investigations of the links between EUV brightenings and the line-of-sight magnetic field suggested that the majority of these events occurred co-spatial with bipoles (see \citealt{Panesar21,Kahil22}). Our research displays a slightly different picture. When using our base criteria ($m=5$ and {\bf{B}}$_\mathrm{th}=20$ G), for example, only $12.4$ \%\ of EUV brightenings are found to form co-spatial with strong bipoles. Even with our loosest criteria ($m=9$ and {\bf{B}}$_\mathrm{th}=20$ G) we found that only $47$ \%\ of EUV brightenings form co-spatial with strong bipoles. One may question whether these differences in results could be explained purely through differences in methodology. Notably, previous studies have analysed manually detected EUV brightenings whereas we analyse automatically detected events. We find that EUV brightenings that are detected co-spatial with strong bipoles are, on average, slightly larger and longer-lived than other events which would make them easier to observe manually, thereby, potentially introducing a slight bias into previous results. However, only $17.9$ \%\ of the largest EUV brightenings (areas above $1$ Mm$^2$, equivalent to $33$ \HRI\ pixels) are found co-spatial with strong bipoles using our base criteria suggesting any biases introduced by this should be small. Increasing the area that was searched over to identify strong bipoles or lowering {\bf{B}}$_\mathrm{th}$ further would increase the percentage of events categorised as such, though this would also weaken the argument that the photospheric magnetic field structuring is linked to the observed coronal dynamics. 

Although we do not find that EUV brightenings predominantly form co-spatial with strong bipoles in the solar photosphere, we do find that EUV brightenings occur predominantly co-spatial with the strong line-of-sight magnetic field that comprises the photospheric network. Depending on the threshold, between $43.4$ \%\ ($m=5$ and {\bf{B}}$_\mathrm{th}=50$ G) and $97.1$ \%\ ($m=9$ and {\bf{B}}$_\mathrm{th}=20$ G) of EUV brightenings occur co-spatial with strong photospheric line-of-sight field, with the true value potentially being close to $80.9$ \% (associated with our base criteria, $m=5$ and {\bf{B}}$_\mathrm{th}=20$ G). Previous research using numerical methods has suggested that EUV brightenings are consistent with the signatures expected due to the relaxation of the magnetic field through a series of non-potential states (\citealt{Barczynski22}), potentially due to the occurrence of component magnetic reconnection (\citealt{Chen21}). Additionally, \citet{Alipour22}, found that around $27$ \%\ of EUV brightenings are found within larger coronal bright point structures presumably over bipolar photospheric network elements. An apparent connection, as observed here, between EUV brightenings and the photospheric network is, therefore, perhaps to be expected. Intuitively, these regions of strong line-of-sight magnetic field are the most obvious candidates to host the free magnetic energy required to allow such relaxation/reconnection in the quiet Sun.

On the other hand, the converse of this result is that we still find a significant number of EUV brightenings ($19.1$ \%\ uing our base criteria of $m=5$ and {\bf{B}}$_\mathrm{th}=20$ G) that occur in regions where no strong line-of-sight magnetic field is measured. Whether the models of \citet{Barczynski22} and \citet{Chen21} could explain events similar to the Weak Field example plotted in Fig.~\ref{CFs_fig3} is currently uncertain. The relaxation/reconnection model would require the presence of either unresolved photospheric line-of-sight magnetic field, or the presence of long (several Mm in length) horizontal field components (with no signature in the line-of-sight componenet) that connect to photospheric foot-points well away from the EUV brightening in question. Neither of these scenarios are impossible, but more research is required to show that they are occurring. It is also possible, that some other mechanism (e.g., upwardly propagating shocks causing local increases in the density/temperature) is driving the increases in intensity observed in the \HRI\ data in these Weak Field locations. A dedicated analysis of the most isolated EUV brightenings, ideally combined with other datasets that clearly show shocked acoustic grains (such as IRIS chromospheric and transition region imaging data; \citealt{Martinez15}), would be required to test this further.

If magnetic reconnection is the driver of some EUV brightenings, then one may expect that flux cancellation would be present at the foot-points of those events that occur over strong bipoles in particular (as is found for other localised brightenings in the solar atmosphere; \citealt{Young18}). Investigating the $627$ EUV brightenings that occurred co-spatial with strong bipoles using our base criteria ($m=5$ and {\bf{B}}$_\mathrm{th}=20$ G), though,  did not return clear evidence that this was the case. Indeed, only $54$ events were found co-spatial with strong cancellation ($<-10^{15}$ Mx s$^{-1}$) in the local photospheric magnetic flux. Interestingly, slightly more events ($57$) were found to occur co-spatial with strong increases in the local magnetic flux ($10^{15}$ Mx s$^{-1}<$), potentially associated with flux emergence. We note that our aim was to conduct a statistical analysis that was consistent across many events and, therefore, our methodology may have contributed to the low number of events associated with cancellation and emergence. For example, we only studied the evolution of the magnetic flux over short time-periods (five SDO/HMI frames) in order to allow those events close to the start or end of the entire time-series to be included. We also only studied small FOVs ($5\times5$ pixel$^2$) to allow events close to the edge of the entire FOV to be included. Future more targeted case studies of a large number of events (ideally hundreds) using data from Solar Orbiter's PHI, or even higher-resolution instruments, could provide further insights into the percentage of events that are truly co-spatial with cancellation or emergence.

\subsection{Links between EUV brightenings and other solar features}

A natural question that arises from these discussions is: What implications do these results have about the potential relationship between EUV brightenings and other localised brightenings observed in different thermal regimes within the solar atmosphere? The single event sampled by the IRIS spectrograph slit in the work of \citet{Nelson23} (who studied this same dataset) had transition region spectral profiles consistent with an Explosive Event (\citealt{Brueckner83}) which are typically modelled by magnetic reconnection between opposite polarity, typically horizontal, field in low-beta plasmas (\citealt{Innes99}). If we consider our four categories then Strong Bipolar and Unipolar field are possibly less likely to provide the magnetic field configurations required to form Explosive Events in the upper solar atmosphere, being (presumably) more vertical in nature. It is possible, therefore, that the Weak Mixing and Weak Field categories could be the most promising for identifying the sub-set of EUV brightenings that are most likely to occur co-spatial with Explosive Events. To initially explore this, we checked the EUV brightening observed by \citet{Nelson23} and found, as expected, that this event was grouped into the Weak Mixing category. Additional studies analysing additional coordinated observations are required to make further progress in this area. 

Regarding Strong Bipolar EUV brightenings, these could also be linked to other localised brightenings in different thermal regimes within the solar atmosphere. Ellerman bombs (EBs; \citealt{Ellerman17}) and IRIS bursts (\citealt{Peter14}), for example, are both typically observed co-spatial with bipoles (often cancelling) in the photospheric line-of-sight magnetic field in active regions (\citealt{Young18}). Although these events have formation temperatures than are separated by around one order of magnitude ($\sim8\times10^3$ K for EBs and $\sim8\times10^4$ K for IRIS bursts), they are often observed to occur co-temporally along the same line-of-sight (\citealt{Vissers15}), which has been successfully modelled by magnetic reconnection occurring at multiple heights along an extended vertical current sheet (\citealt{Hansteen19}).  Given up to $47$ \%\ of EUV brightenings form co-spatial with strong bipoles in the solar photosphere, it is possible that the inclusion of an even more extended current sheet into the EB/IRIS burst model could explain the formation of a significant number of these events. We note that EBs/IRIS bursts and the EUV brightenings studied here occur in very different regions of the solar atmosphere (active regions and the quiet-Sun, respectively), however, the presence of Quiet-Sun Ellerman-like Brightenings (QSEBs; \citealt{Rouppe16,Nelson17}) in the quiet-Sun suggests that similar physics can occur across the entire solar surface. Analysing coordinated datasets sampled by Solar Orbiter and the Swedish Solar Telescope (SST; \citealt{Scharmer03}) would allow an estimation of the percentage of QSEBs that form along the same line-of-sight as EUV brightenings, thereby, providing further insights.

\section{Conclusions}
\label{Conclusions}

The definition of EUV brightenings is relatively simple: Transient increases in intensity, above the noise level, detected in EUV imaging data. EUV brightenings themselves, however, are very complex. Some of these events display signatures in imaging data sampling cooler temperatures, whilst others do not (\citealt{Nelson23}). Some of these events occur within larger scale structures, whilst others do not (\citealt{Alipour22}). Some of these events occur co-spatial with bipoles in photospheric line-of-sight magnetic field maps, whilst others do not (\citealt{Kahil22}). In this work, we have added to this picture by conducting a larger scale statistical investigation of the connections between $5064$ EUV brightenings identified in the quiet-Sun and the line-of-sight magnetic field in the solar photosphere. This sample is two orders of magnitude larger than considered in previous similar studies. Our main conclusions are:
\begin{itemize}
\item{EUV brightenings are preferentially found co-spatial with regions of strong (absolute field strengths above $20$ G) line-of-sight magnetic field in the quiet-Sun photosphere. Around $80.1$ \%\ of EUV brightenings are found co-spatial with strong magnetic field using our base criteria ($m=5$ and {\bf{B}}$_\mathrm{th}=20$ G; see Sect.~\ref{Magnetic}), despite only $53.8$ \%\ of SDO/HMI pixels matching the criteria to be defined as such. When looser or stricter thresholds are considered, the percentage of EUV brightenings co-spatial to strong line-of-sight magnetic fields is found to vary between $43.4$ \%\ and $97.1$ \%. As such, we can consider these values as the lower and upper bounds for the percentage of EUV brightenings that are co-spatial with the photospheric network.}
\item{Most EUV brightenings do not occur co-spatial with strong bipoles. Only $12.4$ \%\ of EUV brightenings are found co-spatial with strong bipoles in the solar photosphere using our base criteria. This percentage rises to $47$ \%\ using our loosest criteria ($m=9$ and {\bf{B}}$_\mathrm{th}=20$ G), however, this is still well below previous estimates reported in the literature.}
\item{A significant number ($19.1$ \%\ using our base criteria) of EUV brightenings are detected in Weak Field regions, where no strong line-of-sight magnetic field is measured. Whether such events can be consistently modelled using current theories about the formation mechanisms responsible for these events remains to be seen. Additionally, fewer than $10$ \%\ of EUV brightenings are detected to occur co-spatial with Unipolar field in the solar atmosphere.}
\item{The number of EUV brightenings that occur co-spatial with strong cancellation or emergence is smaller than may be expected. Indeed, only $54$ EUV brightenings ($1.1$ \%\ of all events) are found to be Strong Bipolar and to display strong evidence of cancellation (rates below $-10^{15}$ Mx s$^{1}$). Additionally, only $57$ EUV brightenings ($1.1$ \%\ of all events) are found to be Strong Bipolar and to display strong evidence of emergence (rates above $10^{15}$ Mx s$^{1}$). The number of EUV brightenings co-spatial with both cancellation and emergence does increase if we consider weaker rates, however, a more targeted study would be better placed to truly test whether these signatures are above noise level.}
\end{itemize}
In order to build on these conclusions going forward, additional studies should be conducted in different locations in the solar atmosphere. One might, for example, expect different statistics in regions that contained extended Unipolar plage. Additionally, other studies using coordinated observations collected by a range of telescopes (e.g., SST, IRIS, PHI, EUI) that sample different observables at various heights in the solar atmosphere may offer further insights into the relationship between these events and other localised brightenings in different thermal regimes. Both MUSE and EUVST will also allow rapid advances in this area in the near future if appropriate datasets can be collected.

\begin{acknowledgements}
This research was supported by the International Space Science Institute (ISSI) in Bern, through ISSI International Team project \#23-586 (‘Novel Insights Into Bursts, Bombs, and Brightenings in the Solar Atmosphere from Solar Orbiter’). CJN, LAH, and SM are thankful to ESA for support as ESA Research Fellows. NF is supported by NASA under contract NNG09FA40C (IRIS) and NNG04EA00C (SDO/AIA). DML is grateful to the Science Technology and Facilities Council for the award of an Ernest Rutherford Fellowship (ST/R003246/1). SP acknowledges the funding by CNES through the MEDOC data and operations center. CV, AZ and DB thank the Belgian Federal Science Policy Office (BELSPO) for the provision of financial support in the framework of the PRODEX Programme of the European Space Agency (ESA) under contract numbers 4000143743, 4000134088, and 4000136424. Solar Orbiter is a mission of international cooperation between ESA and NASA, operated by ESA. The EUI instrument was built by CSL, IAS, MPS, MSSL/UCL, PMOD/WRC, ROB, LCF/IO with funding from the Belgian Federal Science Policy Office (BELSPO/PRODEX PEA 4000134088, 4000112292, and 4000106864); the Centre National d’Etudes Spatiales (CNES); the UK Space Agency (UKSA); the Bundesministerium f\"ur Wirtschaft und Energie (BMWi) through the Deutsches Zentrum f\"ur Luft und Raumfahrt (DLR); and the Swiss Space Office (SSO). SDO/AIA and SDO/HMI data provided courtesy of NASA/SDO and the AIA and HMI science teams. We are thankful to Wei Liu and the AIA science team for facilitating the increased cadence of the instrument during this observing window. This research has made use of NASA’s Astrophysics Data System Bibliographic Services.
\end{acknowledgements}

\bibliographystyle{aa}
\bibliography{SOLO_CFs}

\appendix

\section{Statistical distributions of EUV brightenings}

\begin{table*}[!hbt]
\centering
\caption{Statistics of EUV brightenings related to the line-of-sight magnetic field configuration for eight different detection thresholds.}
\begin{tabular}{ c  c  c  c  c  c  c  c  c  c  c  c  c  c  c  c  c  c  c }
\hline
\multirow{2}{*}{\bf{B}$_\mathrm{th}$} & HMI & & \multicolumn{3}{c }{\color{blue}\bf{Strong Bipolar}} & & \multicolumn{3}{c }{\color{yellow}\bf{Weak Mixing}} & & \multicolumn{3}{c }{\color{green}\bf{Unipolar}} & & \multicolumn{3}{c }{\color{red}\bf{Weak Field}}\\
& FOV & &  Events & $\overline{\mathrm{area}}$ & $\overline{\mathrm{dur}}$ & & Events & $\overline{\mathrm{area}}$ & $\overline{\mathrm{dur}}$ & & Events & $\overline{\mathrm{area}}$ & $\overline{\mathrm{dur}}$ & & Events & $\overline{\mathrm{area}}$ & $\overline{\mathrm{dur}}$ \\
G & pixel$^2$ & & \% & Mm$^2$ & s & & \% & Mm$^2$ & s & & \% & Mm$^2$ & s & & \% & Mm$^2$ & s \\
\hline
20 & 5x5 & & 12.4 & 0.38 & 27 & & 59.3 & 0.31 & 25 & & 9.2 & 0.25 & 27 & & 19.1 & 0.32 & 25 \\
30 & 5x5 & & 5.1 & 0.43 & 31 & & 46.2 & 0.30 & 25 & & 9.2 & 0.25 & 27 & & 39.4 & 0.33 & 25 \\
40 & 5x5 & & 3.1 & 0.45 & 29 & & 37.8 & 0.30 & 25 & & 9.1 & 0.25 & 27 & & 50.0 & 0.33 & 25 \\
50 & 5x5 & & 2.1 & 0.48 & 34 & & 32.2 & 0.30 & 25 & & 9.0 & 0.25 & 27 & & 56.7 & 0.32 & 25 \\
\hline
20 & 9x9 & & 47.0 & 0.35 & 28 & & 49.7 & 0.27 & 24 & & 0.4 & 0.21 & 32 & & 2.9 & 0.36 & 27 \\
30 & 9x9 & & 22.5 & 0.40 & 31 & & 61.5 & 0.28 & 24 & & 0.4 & 0.21 & 32 & & 15.5 & 0.31 & 25 \\
40 & 9x9 & & 15.1 & 0.40 & 30 & & 58.7 & 0.29 & 25 & & 0.4 & 0.21 & 32 & & 25.8 & 0.31 & 24 \\
50 & 9x9 & & 11.0 & 0.39 & 31 & & 56.5 & 0.30 & 26 & & 0.4 & 0.21 & 32 & & 32.1 & 0.31 & 24 \\ 
\hline
\end{tabular}

\tablefoot{Same as Table~\ref{Tab_overview} but for percentages instead of number of events.}
\label{Tab_percentages}
\end{table*}

\begin{figure*}[!hbt]
\includegraphics[width=0.99\textwidth,trim={0 4.2cm 0 0}]{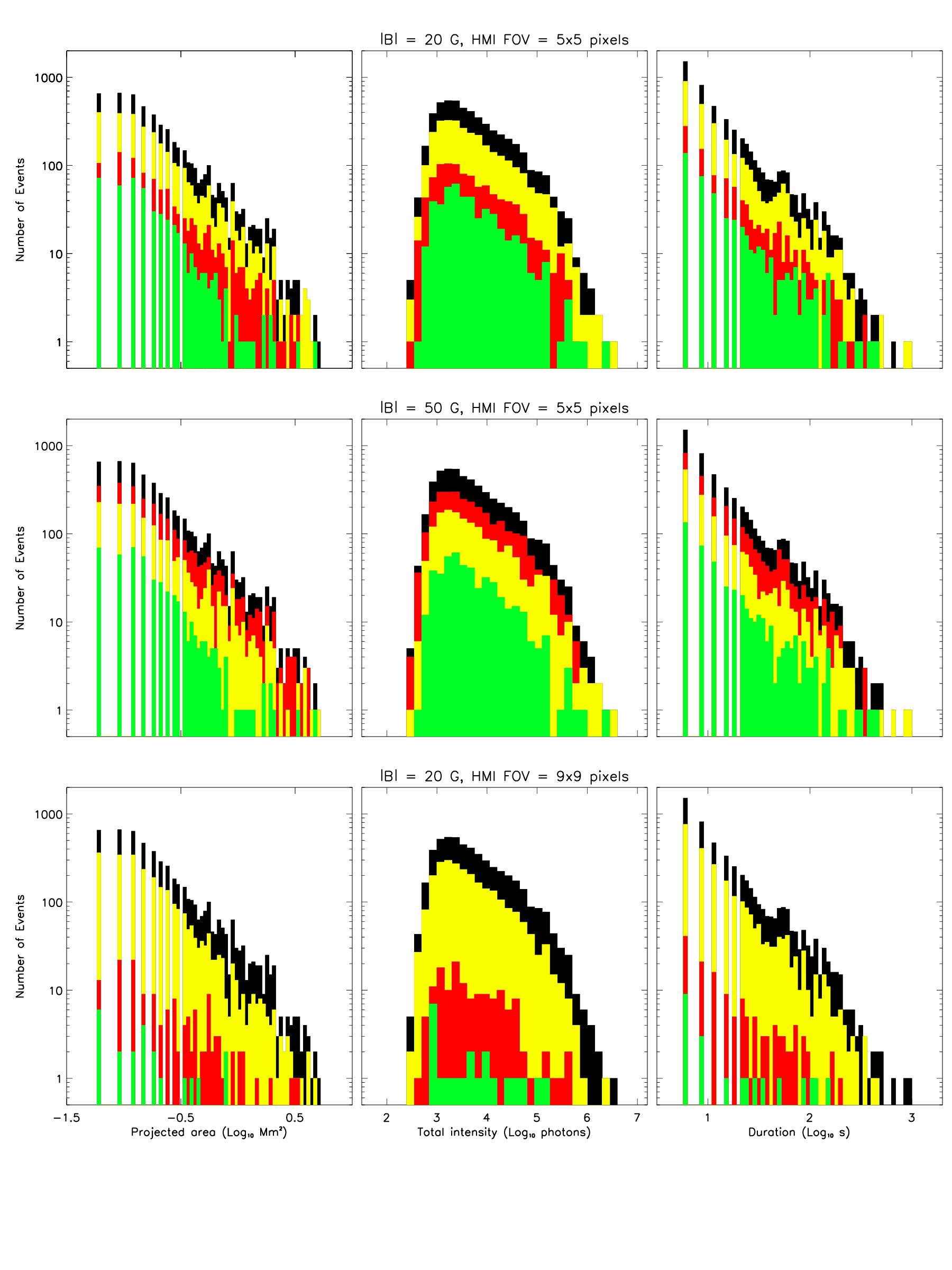}
\caption{These plots are similar to Fig.~\ref{CFs_fig4} except plotting the distributions from the three other categories of EUV brightenings (Weak Mixing, Unipolar, Weak Field) for three different $m$ and {\bf{B}}$_\mathrm{th}$ pairings. The black bars plot the distributions for all EUV brightenings, whilst the other colours are the same as used in Fig.~\ref{CFs_fig3} (yellow for Weak Mixing, red for Weak Field, green for Unipolar). Top row: Statistical distributions for our base pairing of $m=5$ and {\bf{B}}$_\mathrm{th}=20$ G. Middle row: Statistical distributions for our strictest pairing of $m=5$ and {\bf{B}}$_\mathrm{th}=50$ G. Bottom row: Statistical distributions for our loosest pairing of $m=9$ and {\bf{B}}$_\mathrm{th}=20$ G.}
\label{CFs_fig1A}
\end{figure*} 

\end{document}